\documentclass[%
reprint,
 amsmath,amssymb,
 aps,
prb,
]{revtex4-2}

\usepackage{graphicx}
\usepackage{dcolumn}
\usepackage{bm}

\usepackage{amsmath,empheq}
\usepackage{graphicx}
\usepackage[colorlinks=true, allcolors=blue]{hyperref}
\usepackage{braket}
\usepackage{xcolor}
\usepackage[T1]{fontenc}

\begin{document}

\title{Magneto-optical relaxation dynamics of localized excitons in (In)GaSb/AlGaSb quantum wells and dots}

\author{
G. Barbieri$^{1,2}$,
J. M. Llorens$^{1}$,
A. Chellu$^{3}$,
J. Hilska$^{3}$,
T. Hakkarainen$^{3}$,
and B. Al\'en$^{1}$\\[0.5em]
{\small\itshape
$^{1}$Instituto de Micro y Nanotecnolog\'ia, IMN-CNM, CSIC, Calle de Isaac Newton 8, E-28760 Tres Cantos, Spain\\
$^{2}$Departamento de Física de Materiales, Facultad de Ciencias Físicas, Universidad Complutense de Madrid, E-28040 Madrid, Spain\\
$^{3}$Optoelectronics Research Centre, Physics Unit, Korkeakoulunkatu 3, 33720 Tampere, Finland
}
}

\begin{abstract}
Local-droplet-etched (In)GaSb/AlGaSb quantum dots are promising silicon-compatible telecom emitters, but their performance can be influenced by slow carrier relaxation under non-resonant excitation of their surrounding quantum well. We investigate the capture dynamics of the quantum well-dot system using time-, energy- and polarization-resolved photoluminescence under magnetic fields up to $9~\mathrm{T}$. The measurements reveal downhill exciton relaxation through disorder-localized states; a localized-tail rate-equation model reproduces the resulting spectral evolution and energy-dependent decay. Extending this framework to include Zeeman splitting and spin-flip transfer allows us to describe the polarization-resolved dynamics, yielding effective exciton $g$-factors on the quantum well between $-1.07$ and $-1.56$ and spin-flip times of $0.2$-$0.6~\mathrm{ns}$. With kinetic parameters constrained by time-resolved measurements, the model also reproduces the steady-state spectra across different well structures. This combined analysis characterizes the interplay of disorder-mediated relaxation and magnetic-field-dependent spin dynamics in the carrier reservoir preceding quantum-dot emission.
\end{abstract}

\maketitle

\section{Introduction} \label{sec:intro}

Local-droplet etching (LDE) yields GaSb/AlGaSb quantum dots (QDs) with low areal density, good size uniformity, and emission in the telecom S-band \cite{chellu2021highly, michl2023strain, leguay2024unveiling}. Related droplet-epitaxy routes have already delivered low-density site-controlled InAs/GaAs QDs \cite{alonso2007formation, martin2009single} and highly symmetric GaAs/AlGaAs QDs \cite{gurioli2019droplet}, the latter with negligible fine-structure splitting \cite{huo2013ultra}, single-photon purity reaching $g^{(2)}(0) \sim 7 \times 10^{-5}$ \cite{schweickert2018demand}, and entanglement fidelities of up to 0.978 \cite{huber2018covre}. Antimonide heterostructures additionally offer a favorable route toward monolithic integration on silicon \cite{huang2006strain, huang2008simultaneous}, so carrying these advantages over to that material system would open a direct path to silicon-compatible telecom quantum emitters.

LDE-grown GaSb/AlGaSb QDs already exhibit several of the properties required for this goal. The first reported structures showed areal densities of approximately $2.6 \times 10^{7}$~cm$^{-2}$ and an ensemble inhomogeneous linewidth of about 8~meV \cite{chellu2021highly}. Single QD spectroscopy subsequently revealed fine-structure splittings of $12.0 \pm 0.5~\mu$eV and radiative lifetimes of $431 \pm 10$~ps \cite{michl2023strain}. Their electronic structure is strongly influenced by the unusual GaSb/AlGaSb band alignment: 8-band $k \cdot p$ calculations predict a size-dependent indirect-to-direct crossover for small QD heights\cite{leguay2024unveiling}, consistent with the behavior known from two-dimensional GaSb/AlSb heterostructures \cite{cebulla1987direct}. More recently, the incorporation of indium into the active material has extended the emission wavelength towards the telecom C-band \cite{hakkarainen2024telecom}.

The GaSb/AlGaSb band alignment constrains the LDE refill geometry. An active region of GaSb or InGaSb that is too thin confines the electrons strongly along the growth direction, which favors indirect $L$-valley states and worsens the emitting properties. Preserving bright, predominantly direct-gap emission while also reaching the desired telecom wavelength therefore requires several nanometers of active material to be deposited during nanohole filling. In the structures studied here the nominal refill thickness ranges from 4.88 to 6.10 nm. The same deposition that fills the nanoholes consequently forms a quantum well (QW) in the surrounding planar region. The system is thus a connected QD-QW heterostructure rather than a collection of electronically isolated dots and, under non-resonant excitation, photogenerated carriers relax through QW states before reaching the QDs  \cite{syperek2013impact}.

The time-evolution of photo-excited carriers within this connected structure depends critically on the strength and timescale of different processes, including relaxation and recombination processes taking place on the QW itself previously to QW-QD coupling. Earlier work on these structures have revealed a roughly ten-fold increase in QD ground-state PL intensity between 10 and 60 K, attributed to thermally activated transfer from the QW into the QDs across a barrier of approximately 5 meV \cite{leguay2024unveiling}. Single QD time-resolved photoluminescence (TRPL) has also showed a delayed neutral-exciton emission rise, described by QW-to-QD transfer across a low-energy barrier \cite{michl2023strain, masson2026engineering}. Both results point out to slowing down processes occurring while carriers relax their energy from their photogenerated state energy towards the final QD ground state energy.


For non resonant excitation, disorder-mediated relaxation in the surrounding QW material might control this intermediate evolution and thus the QW dynamics shall be investigated in detail to understand the QW-QD coupled system. In III-V heterostructures, QW width and alloy fluctuations produce local variations of the band gap and confinement energy, broadening the optical response and localizing carriers at low temperature \cite{buyanova2003recombination, bastard1984low}. These localization effects are particularly relevant in thin QWs, where even monolayer-scale thickness variations produce appreciable local shifts of the confined transition energy. The resulting disorder leaves complementary steady-state and time-resolved signatures: a Stokes-shifted and asymmetric PL lineshape, together with faster decay at high energy and delayed emission from deeper states as carriers relax downhill through the localized manifold \cite{amand1989exciton, gurioli1991temperature, gourdon1987picosecond, kusano1989extremely}. In GaSb/AlGaSb QWs, sizeable Stokes shifts and linewidths associated with interface and alloy disorder have already been reported \cite{ferrini1999optical, bottazzi2000optical}, whereas the corresponding energy-resolved relaxation dynamics remain largely unexplored. The dynamical problem can be modeled microscopically, by solving for carriers in a realistic spatially fluctuating potential including phonon-assisted scattering \cite{zimmermann1997excitons, filoche2017localization}, or phenomenologically, through effective rate equations in which downhill transfer competes with local population loss \cite{oueslati1988resonant, golub1998low, rubel2006model, jacobsen2025effects}. In this work, we follow the latter approach. This empirical parametrization is therefore an effective description of the optically sampled localized-state distribution and does not allow to distinguish between microscopic disorder sources. Given the low QD areal density in these samples of the order of $2\times10^{-7}~\mathrm{cm^{-2}}$, QW-QD transfer competes with other radiative and non-radiative population-loss mechanisms likely to occur at the localized carrier reservoir.

A magnetic field provides an additional probe for the localized carrier reservoir. Diamagnetic shifts and field-dependent linewidths reflect the lateral extent of the exciton wave function and the characteristic scale of the disorder potential sampled by the electron-hole pairs \cite{nash1989diamagnetism, giugno1995variation, bansal2007magnetic, laurindo2022spin}. At the same time, localization can strongly modify exciton spin relaxation in III-V QWs \cite{munoz1995spin, maialle2000spin, lu2008localized, fokina2010spin, dirnberger2019ultralong}. These effects are particularly relevant in antimonide QWs, where small effective masses and strong spin-orbit coupling produce sizeable Zeeman splittings and strongly field-dependent spin properties \cite{jancu2004giant, karalic2019gate, shibata2020magnetotransport}. Magneto-optical measurements can therefore determine not only how strongly the QW excitons are localized, but also how their spin populations evolve before QD capture.

This work investigates the energy and spin dynamics of localized electron hole pairs photogenerated non-resonantly in the QWs of LDE-grown (In)GaSb/AlGaSb heterostructures, examining how disorder-mediated relaxation and magnetic-field effects contribute to their optical response. We first compare the QW and QD emission dynamics at $15~\mathrm{K}$ to establish the temporal evolution of the QW population relative to QD emission. Energy-, time- and polarization-resolved measurements of a representative QW at $4~\mathrm{K}$ are then used to investigate downhill relaxation through localized states and its evolution under magnetic fields up to $9~\mathrm{T}$. To quantify these observations, we develop a localized-tail rate-equation model that reproduces the measured spectral and temporal evolution. Its extension to two Zeeman-split branches coupled by spin-flip scattering allows us to distinguish the contributions of energy relaxation and spin redistribution to the polarization-resolved dynamics.
\section{Sample and Methods} \label{sec:methods}

\subsection{Sample structures}

Three samples were grown by solid-source molecular beam epitaxy on $n$-GaSb(100) substrates using the local-droplet-etching and refill procedure developed for GaSb/AlGaSb QDs \cite{chellu2021highly, michl2023strain, leguay2024unveiling}. After oxide desorption, a $100$-nm GaSb buffer layer was deposited, followed by a $50$-nm AlAsSb lower confinement barrier and a $100$-nm AlGaSb matrix layer. Local droplet etching was initiated by depositing $1.5$~ML of Al at $395\,^{\circ}\mathrm{C}$ under a reduced Sb flux of $0.067$~ML\,s$^{-1}$. The resulting Al droplets etched the underlying AlGaSb surface for $180$~s, with AFM measurements of nanoholes finding an areal density of approximately $2.6\times10^{7}$~cm$^{-2}$, an average diameter of about $50$~nm, and an average depth of about $12$~nm. The nanoholes were subsequently refilled with GaSb or InGaSb and overgrown with a further $100$-nm AlGaSb matrix layer. The active region was enclosed by a $50$-nm upper AlAsSb barrier and completed with a $15$-nm GaSb cap. Detailed descriptions of the LDE process, nanohole morphology, and structural characterization are given in Ref.~\cite{chellu2021highly}.

The three samples were designed as a comparative set in which the nominal thickness and composition of the active material were varied while preserving the same overall barrier design. Here, the QW thickness denotes the nominal planar-equivalent thickness of GaSb or InGaSb deposited during the refill step. Sample~B will be used here as the reference sample because it has an intermediate GaSb thickness and will be investigated in detail by energy-resolved TRPL. Relative to Sample~B, Sample~C has a larger GaSb thickness, which is expected to reduce the confinement energy of the QW transition. Sample~A combines a smaller nominal thickness with a non-zero nominal In content, so that increased confinement and the reduced alloy band gap act in opposite directions on the QW transition energy. The parameters are summarized in Table~\ref{tab:structural_parameters}.

\begin{table}[h!]
\caption{\label{tab:structural_parameters} Structural parameters of the investigated (In)GaSb/AlGaSb QW samples. The QW thickness denotes the nominal amount of (In)GaSb deposited during the nanohole refill step.}
\begin{ruledtabular}
\begin{tabular}{cccc}
Sample & QW thickness & QW & Barrier Al content \\
       & (ML / nm)    &    & (\%) \\
\hline
A & 16.0 / 4.88 & $\mathrm{In}_{0.03}\mathrm{Ga}_{0.97}\mathrm{Sb}$ & 32.5 \\
B & 18.6 / 5.67 & $\mathrm{GaSb}$ & 31.9 \\
C & 20.0 / 6.10 & $\mathrm{GaSb}$ & 31.3 \\
\end{tabular}
\end{ruledtabular}
\end{table}

\subsection{Photoluminescence measurements} \label{sec:methods}

Samples A-C were first compared at a lattice temperature of $15$ K and at 0 T. Steady-state photoluminescence (PL) spectra were recorded under non-resonant excitation with an unpolarized $532$-nm continuous-wave (CW) laser. Time-resolved PL was excited wit a ps-pulsed $405$-nm laser diode operating at a repetition rate of $40$ MHz, with temporal traces recorded at the maxima of the QW and QD ground-state emission bands in the three samples.

The detailed steady-state and time-resolved magneto-optical measurements in the Faraday configuration were performed at $4$~K. For each sample, magneto-PL spectra were recorded every 0.5 T up to $9$~T using linearly polarized $785$-nm continuous-wave excitation. Additionally, on sample B, for each field, energy-resolved magneto-TRPL measurements were collected every 1 nm using a ps-pulsed $780$-nm laser diode operating at a repetition rate of $40$~MHz.

For all measurements, the emission was collected in back-scattering geometr and spectrally dispersed by a $300$-mm-focal-length monochromator. Steady-state PL was detected using a Peltier-cooled InGaAs photomultiplier (PM), while the time-resolved signal was recorded using the same PM connected to a time tagger. Polarization-resolved measurements at $4$~K were performed in the circular basis. To avoid additional modification of the optical path, the two circularly polarized components were obtained by reversing the magnetic-field direction and are denoted by $\sigma^{+}$ and $\sigma^{-}$.
\section{Results} \label{sec:results}

\subsection{Zero field PL and TRPL}

\begin{figure}[h!]
    \centering
    \includegraphics[width=\linewidth]{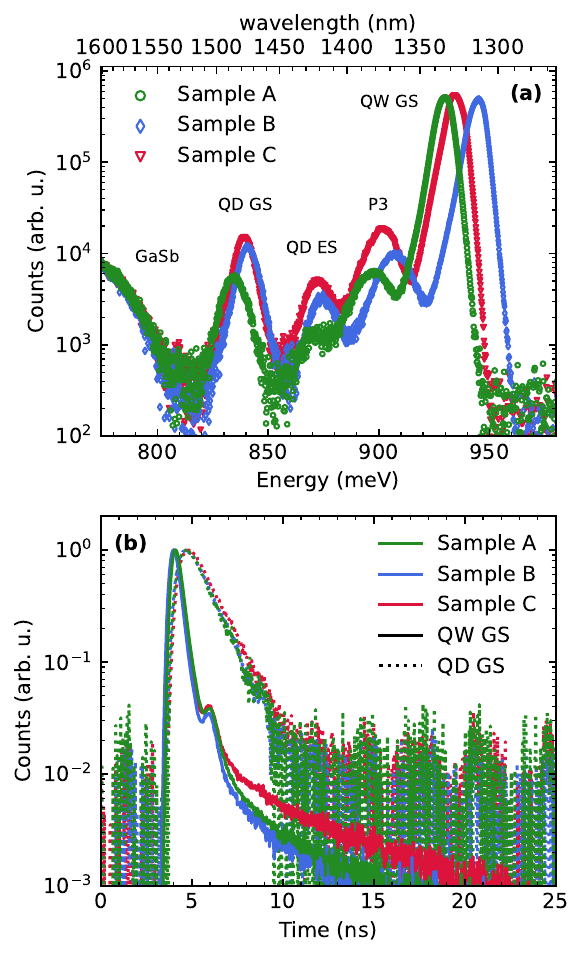}
    \caption{\label{fig:all_spectra} Photoluminescence and time-resolved dynamics of Samples~A-C at $15~\mathrm{K}$. The upper panel shows the full PL spectra, with emission assigned to the GaSb substrate, the QD ground state (QD GS), the QD excited state (QD ES), the additional feature P3, and the QW ground state (QW GS). The lower panel compares the normalized temporal traces recorded at the respective QW and QD GS maxima. }
\end{figure}

Figure~\ref{fig:all_spectra} compares the $15~\mathrm{K}$ PL spectra of Samples~A-C over the full emission range of the samples. From low to high energy, the spectra contain GaSb substrate related emission, the QD ground-state and excited-state ensemble emission, an additional feature denoted P3, which has been attributed to smaller QDs formed near the nanohole rim \cite{chellu2021highly, ulhaq2016vanishing}, and the QW ground-state emission. The same sequence of spectral features is observed in all three samples, consistent with sharing the same architecture despite the variation in refill composition and thickness. The QW maxima occur at approximately $930~\mathrm{meV}$, $946~\mathrm{meV}$, and $936~\mathrm{meV}$ for Samples~A, B, and C, respectively, while the QD ground-state bands lie near $834$-$841~\mathrm{meV}$. The comparatively low QW transition energy of Sample~A, despite its smaller nominal well thickness, is consistent with the reduction of the band gap produced by the In-containing refill layer. In all cases, the QW emission results in an asymmetric lineshape with a low-energy band tail and a peak intensity 35 to 90 times larger than the QD ground state emission of the same sample.

Interface morphology, well-width variations and alloy fluctuations create a spatially varying confinement potential, which is sampled and averaged over by the exciton wave function \cite{buyanova2003recombination, rubel2006model, zimmermann1997excitons}. At $15~\mathrm{K}$, the QW emission has full-width-half maximum (FWHM) values of $8.5$, $8.5$ and $8.8~\mathrm{meV}$ for Samples A, B and C, respectively (Figure~\ref{fig:all_spectra}(a)). Linewidths of approximately $8~\mathrm{meV}$ have been reported for GaSb/AlGaSb QWs of similar thickness and attributed to interface and alloy disorder \cite{ferrini1999optical}, and larger broadenings for GaSb/AlGaSb multiple-QW structures affected in addition by well-width fluctuations and strain relaxation \cite{bottazzi2000optical}. For comparison, low-temperature exciton PL linewidths reported for GaAs/AlGaAs QWs range from about $0.4~\mathrm{meV}$ in high-quality $12~\mathrm{nm}$ wells \cite{ubyivovk2003spectroscopy} to about $8~\mathrm{meV}$ in narrower or more disorder-affected wells \cite{haldar2017effect}.

The asymmetric lineshape alone does not establish redistribution between localized states, and needs of energy-resolved dynamics to provide a direct evidence for downhill relaxation through this manifold. The lower panel of Figure~\ref{fig:all_spectra} shows in log-scale the normalized TRPL decay curves recorded at 15 K for the QW and QD ground-states. The results are similar in all three samples, with small differences only visible for the long-lived QW emission beyond 6 ns. The QW fast rise and decay traces is in clear contrast with the QD ground-state emission which rises and decays more slowly, the peak response of the QD being delayed with respect to the laser excitation by around 600 ps. 


This temporal hierarchy is consistent with the slow QD luminescence rise and thermally activated QW-to-QD transfer reported previously for these structures \cite{michl2023strain, leguay2024unveiling,masson2026engineering}, and places them in the slow-transfer regime recently distinguished from structures exhibiting ultrafast QW-QD coupling \cite{masson2026engineering}. Whether this QD delayed response is related or not with the $\Gamma-L$ crossover in the GaSb/AlGaSb system is still controversial. As we shall see in the following, a careful analysis of the spectrally and time resolved QW photoluminescence reveals that carriers undergo substantial hopping through the non-equilibrium landscape of disorder-localized energy states before being transferred to the QDs in around 400 ps. This time is smaller but comparable to the measured QD radiative lifetime (~1 ns) and leads to  secondary carrier capture and multiphoton emission when exciting non-resonantly the QDs.\cite{masson2026engineering} According to our observations, direct relaxation paths through this level structure hierachy accounts for 96\% of the observed QW+QD luminescence at low temperatures. Carrier capture in and release from a moderately coupled L-valley could still be related to the long-lived decay component observed in the QW, but its impact in the overall dynamics is indistinguishable in our experiments from other intrinsic and extrinsic mechanisms.



\subsection{Magnetic dependence}

\begin{figure}[h!]
    \centering
    \includegraphics[width=\linewidth]{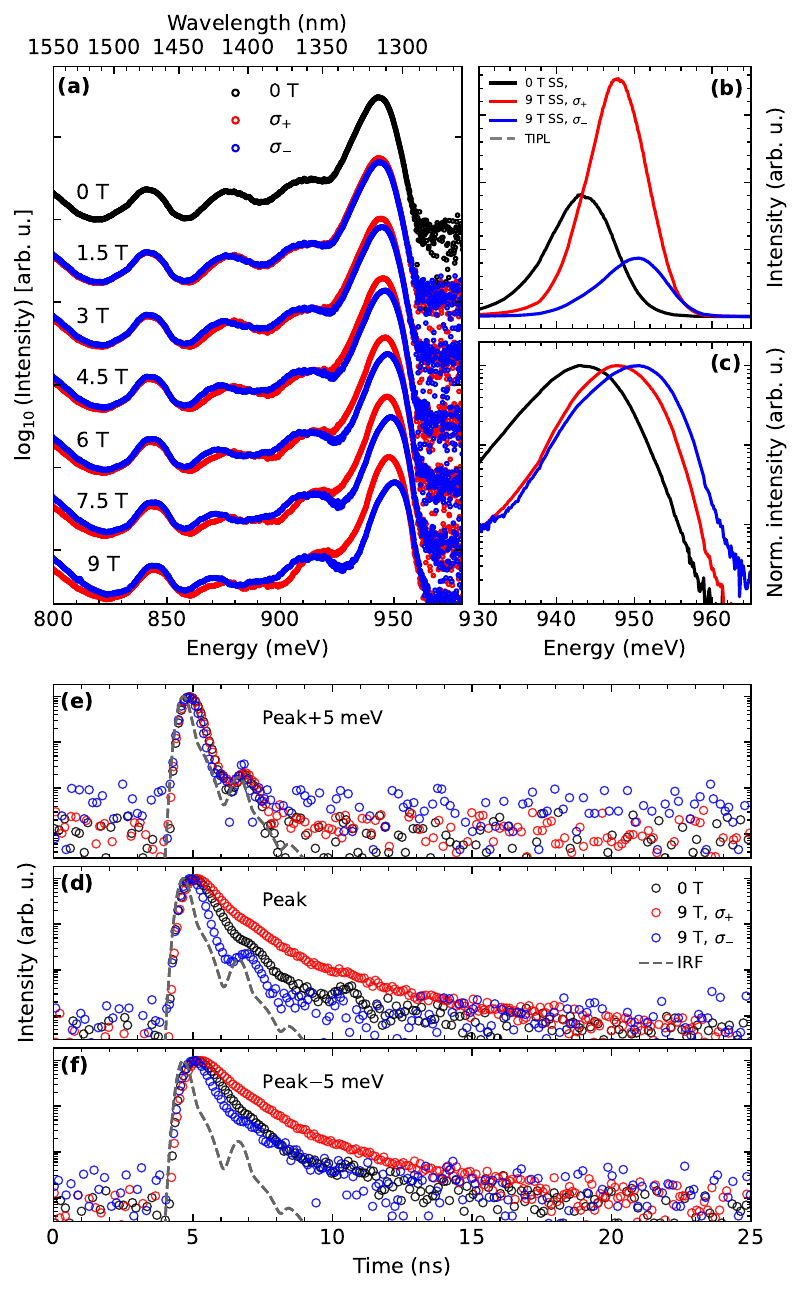}
    \caption{\label{fig:4K_spectra}
    Spectral and temporal QW emission of Sample~B at $4~\mathrm{K}$. (a) Polarization-resolved PL spectra between $0$ and $9~\mathrm{T}$, vertically offset for clarity. (b) QW emission at $0$ and $9~\mathrm{T}$ (c) Corresponding normalized spectra. (d, e, f) TRPL traces recorded at the emission maxima for varying energy around the PL peak. Circular polarizations and magnetic fields used are shown in the legend. The dashed curves indicate the instrument-response function.
}
\end{figure}

A magnetic field applied perpendicularly to the QW interface shrinks the lateral extent of the exciton wavefunction and potentially its hopping dynamics. Therefore, it can be used to probe the orbital and spin properties in a varying disordered potential landscape \cite{nash1989diamagnetism, erdmann2006diamagnetic, maialle2000spin, laurindo2022spin}. Figure~\ref{fig:4K_spectra} summarizes the steady-state and time-resolved magneto-PL of Sample B at $4~\mathrm{K}$. Figure~\ref{fig:4K_spectra}(a) shows the polarization-resolved PL spectra between $0$ and $9~\mathrm{T}$. At the maximum magnetic field the QW peak shows a substantial degree of circular polarization (four times more integrated intensity in the $\sigma^{+}$ component) as shown in Figure~\ref{fig:4K_spectra}(b). Normalizing the spectra in Figure~\ref{fig:4K_spectra}(c) reveals the diamagnetic shift $5.8~\mathrm{meV}$ and Zeeman splitting $-2.6~\mathrm{meV}$ for the QW emission. Comparatively, the QD bands show a weaker evolution in magnetic field and will be described elsewhere. 


In Figure~\ref{fig:4K_spectra} we also compare normalized temporal traces recorded for different magnetic fields at the corresponding PL peak maxima and at energies $5~\mathrm{meV}$ above and below them. At 0 T, the energy modulation of the decay times can be clearly appreciated. On the high energy tail, the decay transients shown in Figure~\ref{fig:4K_spectra}(d) are barely distinguishable from the impulse response function (IRF) of the system before deconvolution. Moving towards the peak energy and beyond, shown in Figure~\ref{fig:4K_spectra}(e) and (f), the signal clearly slows down. Applying a magnetic field, the dynamics for the spin up and down populations can be also energy resolved. In this case, the $\sigma^{-}$ signal decays clearly faster than the $\sigma^{+}$ component, being the difference more pronounced at the PL peak energy (Figure~\ref{fig:4K_spectra}(e)).

\subsection{Magneto-dependent rate-equations model} \label{sec:model}

To quantify the observed spin and charge dynamics ocurring within the localized exciton reservoir, we formulate a kinetic rate-equations model. Following the framework introduced by Oueslati \textit{et al.} and related approaches \cite{oueslati1988resonant, gourdon1989exciton, mair2000time, su2017generalized}, we describe downhill transfer toward deeper localized states competing with local population loss (radiative recombination and QW-QD transfer). We adapt this framework to the present nonresonantly excited QW by including carrier feeding from higher-energy reservoirs.

For an ideal two-dimensional QW, the density of states has a step-like onset at the lowest QW transition $E_{\mathrm{g}}$. Disorder broadens this edge, introducing an approximately exponential Urbach tail toward lower energies, consistent with theoretical treatments of disorder-induced potential fluctuations \cite{john1986theory, bacalis1988simple, silver1989origin}. We characterize the resulting DOS by a sigmoidal function connecting this tail to the constant two-dimensional DOS at higher energies \cite{cingolani1989electronic, jacobson2005increasing, abramkin2018heterostructures, jacobsen2025effects}. We therefore write
\begin{equation}
\rho(E)=\frac{\rho_{\mathrm{max}}}{1+\exp[-(E-E_{\mathrm{g}})/E_0]} .
\label{eq:dos}
\end{equation}
Here, $E_0$ is the Urbach energy effectively parametrizing the disorder. For $E\ll E_{\mathrm{g}}$, Eq.~\eqref{eq:dos} reduces to $\rho(E)\propto\exp[(E-E_{\mathrm{g}})/E_0]$, whereas for $E\gg E_{\mathrm{g}}$ it approaches $\rho_{\mathrm{max}}$. Since only relative state densities enter the transfer probabilities, the absolute value of $\rho_{\mathrm{max}}$ is not relevant.

\begin{figure}[h!]
    \centering
    \includegraphics[width=\linewidth]{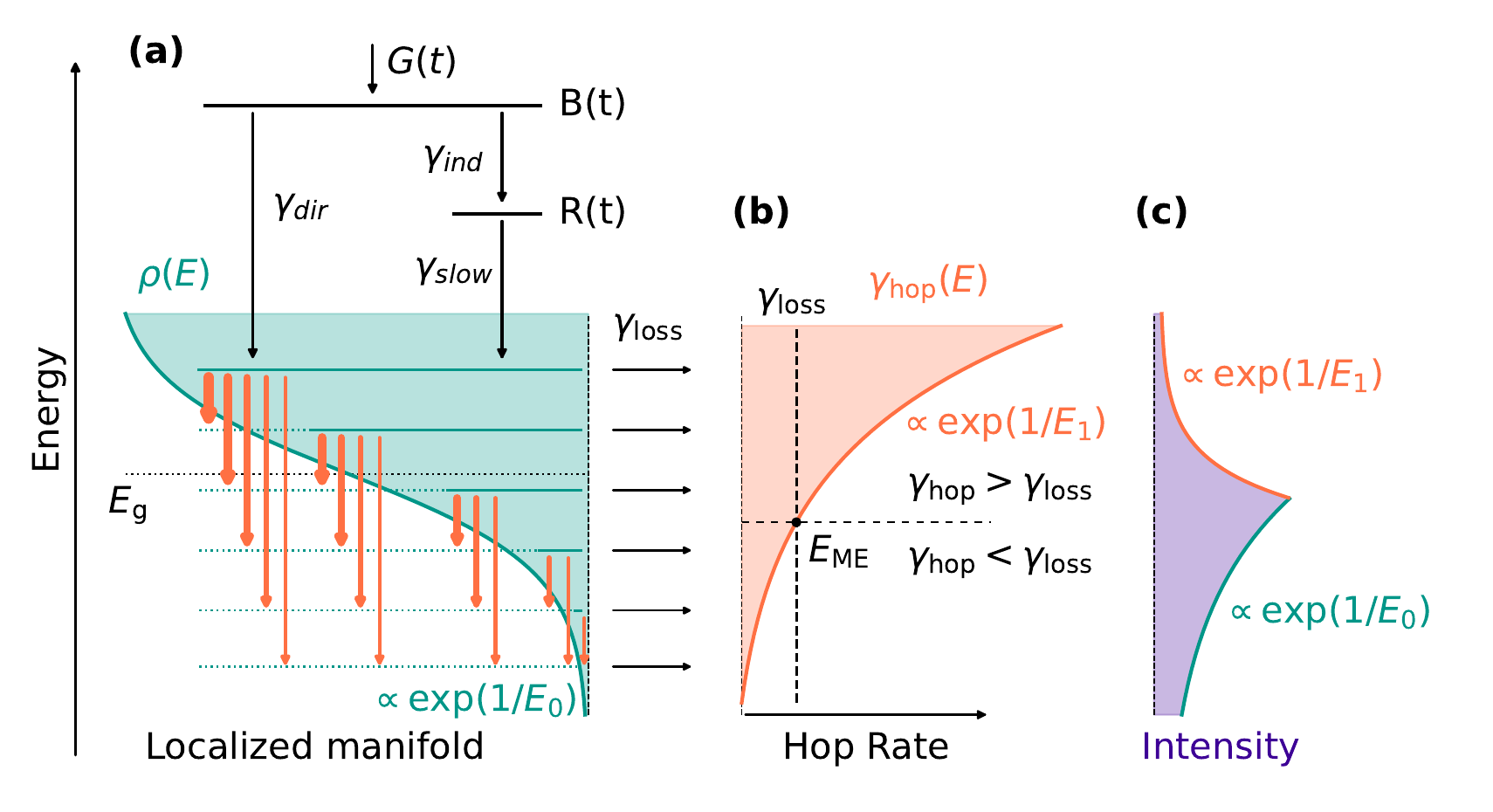}
    \caption{Schematic representation of the localized-tail rate-equation model.
    (a) The QW states follow the disorder-broadened density of states $\rho(E)$, characterized by the effective QW edge energy $E_g$ and the Urbach energy $E_0$. Non-resonantly generated carriers populate the barrier reservoir $B(t)$ and feed the QW either directly at rate $\gamma_{\mathrm{dir}}$ or through the intermediate reservoir $R(t)$ at rates $\gamma_{\mathrm{ind}}$ and $\gamma_{\mathrm{slow}}$. Within the QW manifold, excitons undergo downhill redistribution while being removed locally at rate $\gamma_{\mathrm{loss}}$.
    (b) The hopping rate $\gamma_{\mathrm{hop}}(E)$ decreases toward lower energies with characteristic scale $E_1$. The crossover energy $E_{\mathrm{ME}}$, defined by $\gamma_{\mathrm{hop}}(E_{\mathrm{ME}})=\gamma_{\mathrm{loss}}$, separates transfer- and loss-dominated regimes.
    (c) Illustrative steady-state emission profile resulting from the interplay between the density of available states and downhill redistribution, with the low- and high-energy sides primarily sensitive to $E_0$ and $E_1$, respectively.}
    \label{fig:Schematics}
\end{figure}

The rate-equations model is defined on an energy grid $E_i$ spanning the QW emission window, with $n_i(t)\equiv n(E_i,t)$ denoting the population in energy bin $E_i$. The highest-energy bin of the modeled window is labeled $E_N$. The feeding and relaxation pathways are summarized schematically in Figure~\ref{fig:Schematics}(a). Non-resonant excitation is represented by a generation pulse $G(t)$ that populates an effective barrier reservoir $B(t)$. Carriers feed the QW either directly, at rate $\gamma_{\mathrm{dir}}$, or indirectly through an intermediate reservoir $R(t)$. The latter is introduced to reproduce the non-monoexponential decay observed in the QW transients, with more details left for section \ref{sec:discussion}. The indirect pathway is described by transfer from $B(t)$ to $R(t)$ at rate $\gamma_{\mathrm{ind}}$, followed by delayed release into the QW at rate $\gamma_{\mathrm{slow}}$:
\begin{align}
\dot{B}(t) &= G(t)-\left(\gamma_{\mathrm{dir}}+\gamma_{\mathrm{ind}}\right)B(t),\\
\dot{R}(t) &= \gamma_{\mathrm{ind}}B(t)-\gamma_{\mathrm{slow}}R(t).
\end{align}
The total feeding rate into the QW is therefore
\begin{equation}
S(t)=\gamma_{\mathrm{dir}}B(t)+\gamma_{\mathrm{slow}}R(t).
\end{equation}

The QW population evolves through feeding, local loss, and downhill redistribution:
\begin{align}
\dot{n}_i
&=S(t)\delta_{iN}
-\left[\gamma_{\mathrm{loss}}+\gamma_{\mathrm{hop}}(E_i)\right]n_i \nonumber\\
&\quad+\sum_{j>i}\gamma_{\mathrm{hop}}(E_j)P_{j\rightarrow i}n_j .
\label{eq:rate_qw}
\end{align}
Here, $\delta_{iN}$ represents injection into the highest-energy bin of the modeled QW window. The rate $\gamma_{\mathrm{loss}}$ is assumed to be energy independent and represents removal from the resolved QW manifold. The branching probability for a hop from $E_j$ to a lower-energy bin $E_i$ is taken to be proportional to the density of available final states:
\begin{equation}
P_{j\rightarrow i}=
\frac{\rho(E_i)}{\sum_{k<j}\rho(E_k)}
\qquad (i<j),
\label{eq:transfer_probability}
\end{equation}
such that $\sum_{i<j}P_{j\rightarrow i}=1$.

Following the kinetic picture of Oueslati \textit{et al.} \cite{oueslati1988resonant}, downhill redistribution within the disorder-broadened QW manifold is described by
\begin{equation}
\gamma_{\mathrm{hop}}(E)= \gamma_{\mathrm{loss}} \exp\left(\frac{E-E_{\mathrm{ME}}}{E_1}\right),
\label{eq:outflow}
\end{equation}
which decreases exponentially toward lower energies. The energy $E_{\mathrm{ME}}$ is defined by $\gamma_{\mathrm{hop}}(E_{\mathrm{ME}})=\gamma_{\mathrm{loss}}$ and therefore marks the kinetic crossover between transfer- and loss-dominated regimes. Above $E_{\mathrm{ME}}$, downhill transfer dominates over local removal, whereas below it local loss becomes dominant. $E_{\mathrm{ME}}$ may lie above of below $E_g$. The parameter $E_1$ sets the logarithmic energy dependence of the hopping rate. Its exponential form reflecting the decreasing spatial overlap between the wave functions of progressively deeper localized states \cite{zimmermann1997excitons}.

After finding the populations over energy interval considered, the calculated PL intensity are proportional to:
\begin{equation}
I(E_i,t)=A\,n_i(t),
\end{equation}
where $A$ is a proportionality constant. The resulting spectral profile is illustrated schematically in Figure~\ref{fig:Schematics}(c). The corresponding steady-state spectra are obtained from the stationary populations under continuous feeding. For comparison with the TRPL measurements, the calculated temporal response is convolved with the experimental instrument-response function (IRF).

\begin{figure}[h!]
    \centering
    \includegraphics[width=\linewidth]{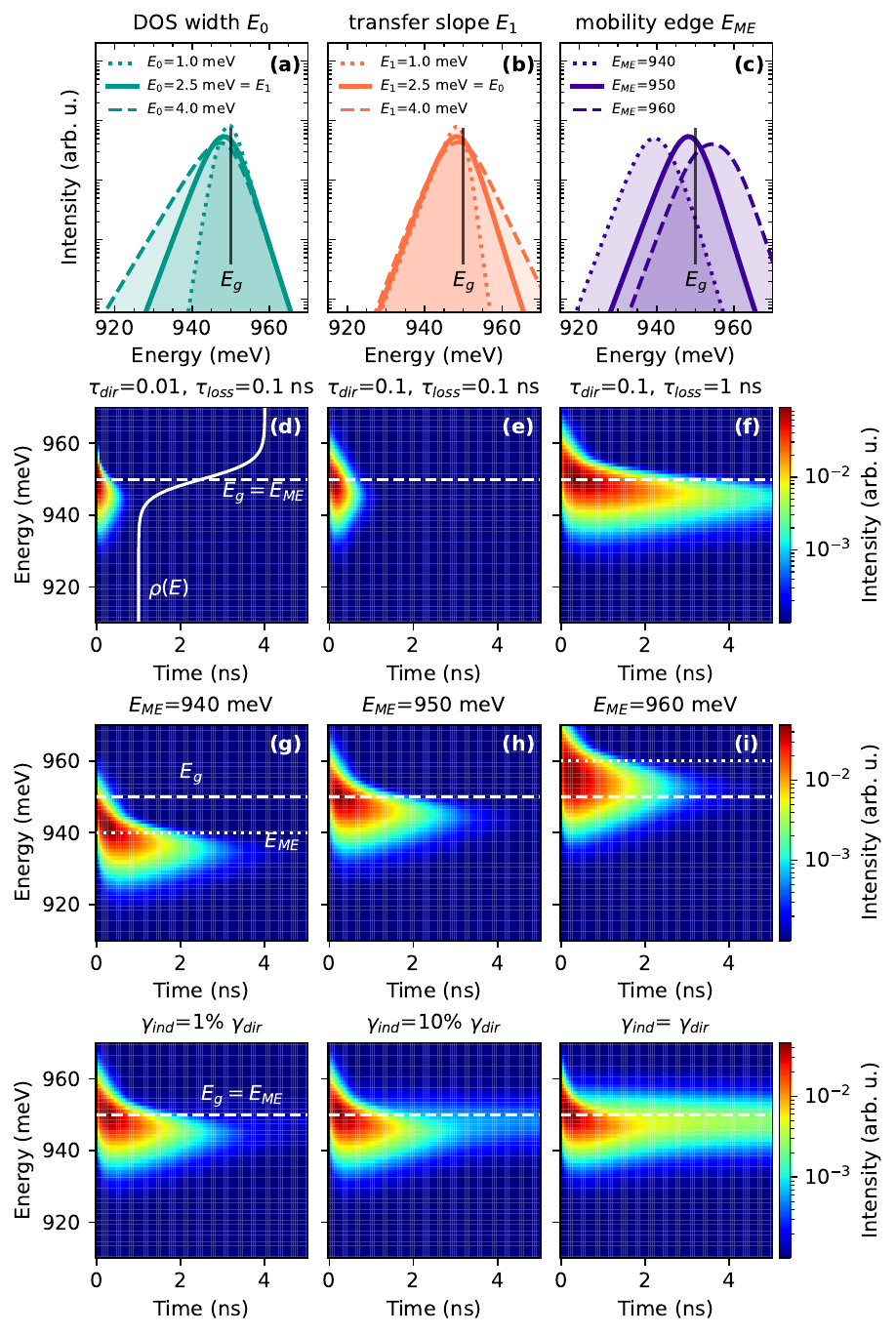}
    \caption{\label{fig:model_response} Illustrative responses of the localized-tail rate-equation model. The steady-state profiles show the influence of the Urbach energy $E_0$ in (a), the hopping-rate energy scale $E_1$ in (b), and the crossover energy $E_{\mathrm{ME}}$ in (c). The energy-time maps illustrate the effects of $\tau_{\mathrm{dir}}$ and $\tau_{\mathrm{loss}}$ in (d)-(f), of varying $E_{\mathrm{ME}}$ relative to the effective QW edge energy $E_g$ in (g)-(i), and of increasing the indirect-feeding contribution in (j)-(l). Dashed and dotted horizontal lines indicate $E_g$ and $E_{\mathrm{ME}}$, respectively, while the white curve in (d) shows the DOS $\rho(E)$. Parameters varied in each panel are indicated in the figure; all remaining parameters are kept fixed.}
\end{figure}

Figure~\ref{fig:model_response} illustrates how the main model parameters shape the calculated emission. Increasing the Urbach energy $E_0$ broadens the low-energy side of the steady-state spectrum (Figure~\ref{fig:model_response}(a)), whereas $E_1$ controls the high-energy slope through the energy dependence of the downhill-transfer rate (Figure~\ref{fig:model_response}(b)). The role of $E_{\mathrm{ME}}$ depends on its position relative to the QW gap $E_g$ (Figure~\ref{fig:model_response}(c)). When $E_{\mathrm{ME}}<E_g$, the population relaxes through the Urbach tail until hopping and local loss become comparable, so that the emission maximum lies close to $E_{\mathrm{ME}}$. When $E_{\mathrm{ME}}>E_g$, the DOS is saturating towards its constant two-dimensional value before this kinetic crossover is reached, resulting in a broader emission profile extending between $E_g$ and $E_{\mathrm{ME}}$.

The energy-time maps separate the effects of the temporal parameters. Increasing the direct-feeding time $\tau_{\mathrm{dir}}=\gamma_{\mathrm{dir}}^{-1}$ delays the initial population of the QW (Figs.~\ref{fig:model_response}(d) and \ref{fig:model_response}(e)), while increasing the local-loss time $\tau_{\mathrm{loss}}=\gamma_{\mathrm{loss}}^{-1}$ prolongs the population at later delays (Figure~\ref{fig:model_response}(f)). The same $E_{\mathrm{ME}}$ vs $E_g$ crossover behaviour is visible dynamically in Figs.~\ref{fig:model_response}(g)-\ref{fig:model_response}(i): for $E_{\mathrm{ME}}<E_g$, the emission relaxes toward energies close to $E_{\mathrm{ME}}$, whereas for $E_{\mathrm{ME}}>E_g$ the populated region spreads across the broader interval between $E_g$ and $E_{\mathrm{ME}}$. Finally, increasing $\gamma_{\mathrm{ind}}/\gamma_{\mathrm{dir}}$ enhances the delayed-feeding contribution, extending the QW dynamics to longer times (Figs.~\ref{fig:model_response}(j)-\ref{fig:model_response}(l)).

For polarization-resolved magneto-PL and magneto-TRPL, the same localized-state kinetics are extended to two branches corresponding to the circularly polarized optical transitions. At magnetic field $B$, a state with underlying zero-field energy $E_i$ appears in branch $s=\pm1$ at the photon energy
\begin{equation}
E_{i,s}(B)=E_i+\Delta_{\mathrm{D}}(B)+s\frac{\Delta_{\mathrm{Z}}(B)}{2},
\label{eq:branch_energy}
\end{equation}
where $\Delta_{\mathrm{D}}(B)$ and $\Delta_{\mathrm{Z}}(B)$ are the diamagnetic shift and Zeeman splitting, respectively. Spin relaxation is introduced by coupling states with the same underlying localization energy $E_i$ and localized-state index. Thus, beyond the energy shifts introduced above, the model only adds a single kinetic parameter, the effective spin-flip rate $\gamma_{\mathrm{s}}$ between the two spin branches. The resulting rate equation for each branch labeled by $s$ is:
\begin{align}
\dot{n}_{i,s}
&=\frac{S(t)}{2}\delta_{iN}
-\left[\gamma_{\mathrm{loss}}+\gamma_{\mathrm{hop}}^\mathrm{s}(E_i)+\gamma_{\mathrm{s}}^{s\rightarrow -s}\right]n_{i,s} \nonumber\\
&\quad+\sum_{j>i}\gamma_{\mathrm{hop}}^\mathrm{s}(E_j)P_{j\rightarrow i}^\mathrm{s}n_{j,s}
+\gamma_{\mathrm{s}}^{-s\rightarrow s}n_{i,-s}.
\label{eq:rate_qw_spin}
\end{align}
The feeding is taken to be spin balanced in this model, because the excitation is non-resonant and linearly polarized. Since the Zeeman splitting can be comparable to $k_{\mathrm{B}}T$, detailed balance is imposed to the transition from the lower- to the higher-energy branch at a rate:
\begin{equation}
\gamma_{\mathrm{s}}
\exp\left[-\frac{|\Delta_{\mathrm{Z}}(B)|}{k_{\mathrm{B}}T}\right].
\end{equation}
The calculated circularly polarized PL signals are finally obtained as
\begin{equation}
I_s(E_{i,s},t)=A\,n_{i,s}(t),
\end{equation}
using the same amplitude factor $A$ for both polarization branches.

\subsection{Parameter extraction}

\begin{figure}[h!]
    \centering
    \includegraphics[width=\linewidth]{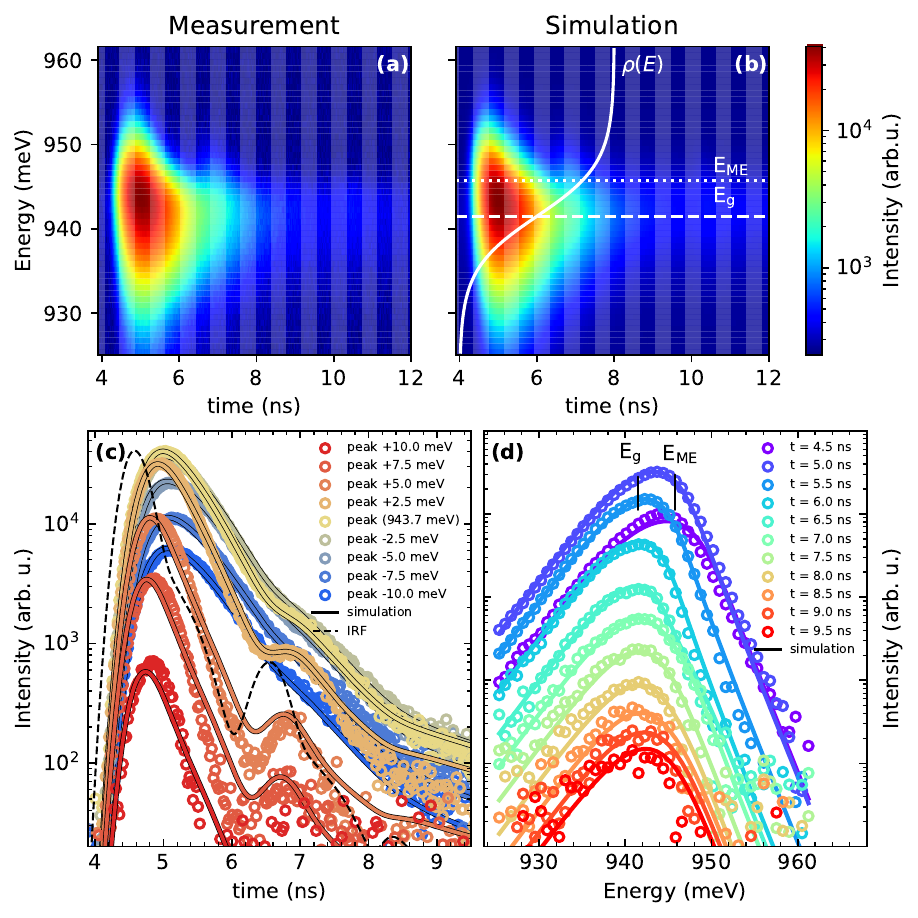}
    \caption{\label{fig:trpl_fit} (a) Measured and (b) simulated energy-resolved TRPL maps. The white curve in (b) shows the fitted DOS $\rho(E)$, while the dashed and dotted lines indicate $E_g$ and $E_{\mathrm{ME}}$, respectively.
    (c) Temporal cuts at selected detection energies; symbols show the measurements, solid curves the simulation, and the dashed curve the IRF.
    (d) Spectral cuts at selected delays, with symbols and solid curves showing the measured and simulated spectra, respectively. The vertical lines indicate $E_g$ and $E_{\mathrm{ME}}$.}
\end{figure}

The energy dependent TRPL of Sample~B at 0 T provides the anchor dataset for extracting the field independent parameters of the rate equations model. The complete energy-time map was fitted globally with a single parameter set, as shown in Figure~\ref{fig:trpl_fit}. Shortly after excitation, the measured QW emission extends over a broad energy interval, with a maximum near $944~\mathrm{meV}$ and a pronounced high-energy contribution extending beyond $950~\mathrm{meV}$. As the population evolves, this high-energy contribution is depleted first, while emission around and below the spectral maximum persists, producing a progressive redshift and narrowing of the QW band. The simulation in Figure~\ref{fig:trpl_fit}(b) reproduces this overall evolution. The fitted parameters are summarized in Table~\ref{tab:TRPL_parameters}. The temporal and spectral cuts in Figs.~\ref{fig:trpl_fit}(c) and \ref{fig:trpl_fit}(d) provide a more direct comparison. The simulated temporal traces span detection energies up to $10~\mathrm{meV}$ above and below the QW maximum, and reproduce nicely the recorded decay traces for all energies. The simulated spectral cuts likewise capture the preferential depletion of the high-energy side and the resulting asymmetric redshift of the emission profile over most of the measured dynamic range.

\begin{table}[h!]
\caption{\label{tab:TRPL_parameters}Parameters obtained from the global fit of the zero-field QW TRPL of Sample~B.}
\begin{ruledtabular}
\begin{tabular}{lc}
Parameter & Value \\
\hline
$E_g$ & $941.5\pm0.1~\mathrm{meV}$ \\
$E_{\mathrm{ME}}$ & $945.7\pm0.1~\mathrm{meV}$ \\
$E_0$ & $3.1\pm0.1~\mathrm{meV}$ \\
$E_1$ & $1.9\pm0.1~\mathrm{meV}$ \\
$\tau_{\mathrm{dir}}$ & $0.22\pm0.01~\mathrm{ns}$ \\
$\tau_{\mathrm{loss}}$ & $0.39\pm0.01~\mathrm{ns}$ \\
$\gamma_{\mathrm{ind}}^{-1}$ & $5.5\pm0.1~\mathrm{ns}$ \\
$\gamma_{\mathrm{slow}}^{-1}$ & $8.5\pm0.1~\mathrm{ns}$ \\
\end{tabular}
\end{ruledtabular}
\end{table}

\begin{figure}[h!]
    \centering
    \includegraphics[width=0.99\linewidth]{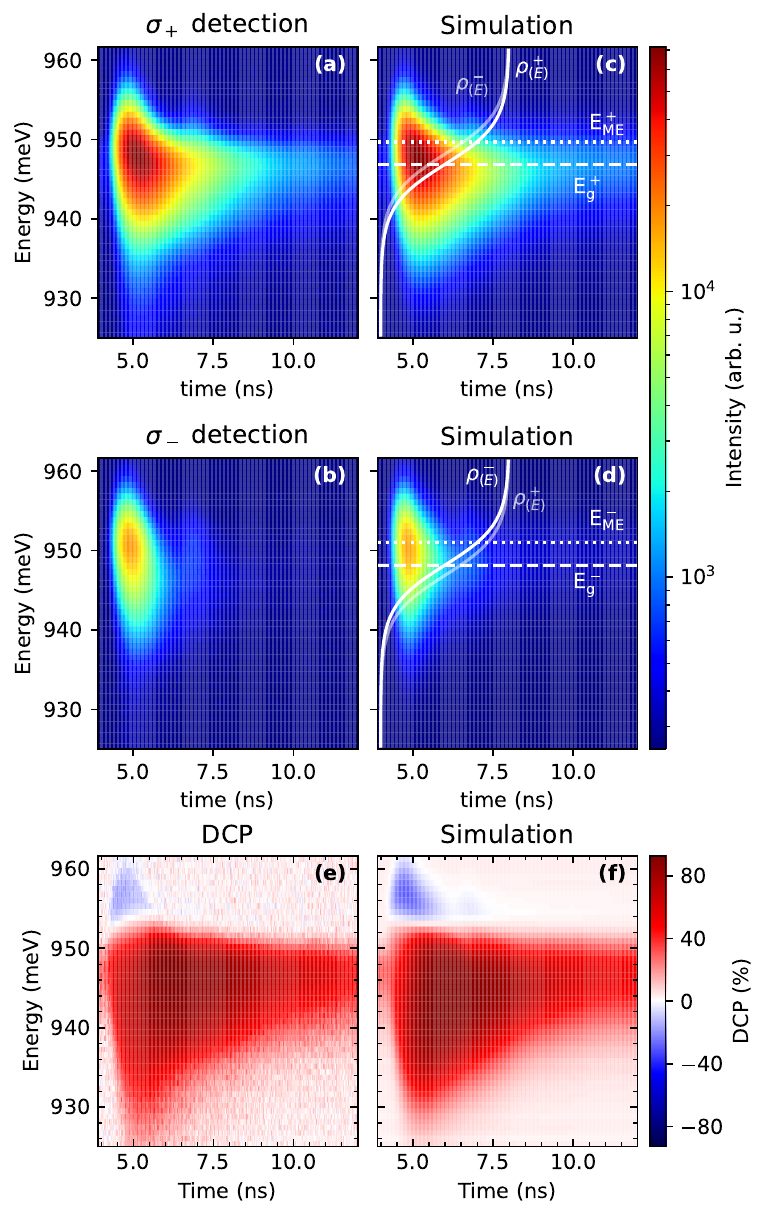}
    \caption{\label{fig:trpl_magneto_fit} Polarization-resolved fit of the QW TRPL of Sample~B at $9~\mathrm{T}$. 
(a),(b) Measured TRPL maps for $\sigma^{+}$ and $\sigma^{-}$ detection, respectively. 
(c),(d) Corresponding simulations obtained from the Zeeman-split extension of the localized-tail model. The white and gray curves show the DOS of the two Zeeman branches, while the dashed and dotted lines indicate the branch-dependen $E_g^{\pm}$ and $E_{\mathrm{ME}}^{\pm}$, respectively. 
(e),(f) Measured and simulated degree of circular polarization.}
\end{figure}

After finding the dynamical parameters of the model at zero field, the polarization-resolved TRPL can be fitted simultaneously for the two Zeeman branches up to 9 T. This results in diamagnetic shift $\Delta_{\mathrm{D}}$, Zeeman splitting $\Delta_{\mathrm{Z}}$, and interbranch spin-flip time $\tau_s$ for each magnetic field. The hopping rate at the effective QW edge is fixed to the zero-field value, $\gamma_{\mathrm{hop}}(E_g)=0.27~\mathrm{ns^{-1}}$. Other DOS and intrabranch kinetic parameters are field dependent but common to both branches. Figure~\ref{fig:trpl_magneto_fit} shows the $9~\mathrm{T}$ dataset, for which the Zeeman splitting and polarization imbalance are most pronounced. The measured $\sigma^{+}$ map in panel~(a) displays a broad initial distribution followed by a long-lived population concentrated toward lower energies, whereas the $\sigma^{-}$ emission in panel~(b) is weaker, spectrally narrower, and confined mainly to the first few nanoseconds. The corresponding simulations in panels~(c) and (d) reproduce the energy separation and pronounced intensity imbalance of the two branches, together with their different temporal evolution. Within each branch, the model retains the rapid depletion of the high-energy states and the progressively slower dynamics toward lower energy. The largest discrepancy occurs in the $\sigma^{-}$ component, for which the simulation underestimates the persistence of the low-energy emission. 
The time- and energy-dependent degree of circular polarization was calculated as
\begin{equation}
\mathrm{DCP}(E,t)=
\frac{I_{\sigma^{+}}(E,t)-I_{\sigma^{-}}(E,t)}
{I_{\sigma^{+}}(E,t)+I_{\sigma^{-}}(E,t)} .
\label{eq:dcp}
\end{equation}
As shown in Figure~\ref{fig:trpl_magneto_fit}(e), the measured DCP is strongly positive over the central and low-energy portions of the QW band, reaching values above $80\%$ where the $\sigma^{+}$ emission dominates. Toward higher energy, the relative contribution of the $\sigma^{-}$ branch increases, causing the DCP to decrease and become weakly negative. The polarization profile also evolves during the decay, with the positive contribution remaining pronounced around the long-lived lower-energy population.

The resulting field-dependent parametrization is summarized in Figure~\ref{fig:param_b} and discussed in the following section.
\section{Discussion} \label{sec:discussion}

The rate-equations model provides an effective parametrization of the observed relaxation-recombination dynamics. The same kinetic framework reproduces the early high-energy response, its rapid depletion, the delayed low-energy maximum, the progressively slower decay toward lower energy, and the evolving spectral profile. The zero-field fit establishes the hopping-rate reference $\gamma_{\mathrm{hop}}(E_g)=0.27~\mathrm{ns^{-1}}$, which is kept fixed in the subsequent field-dependent TRPL and steady-state analyses. Under the same set of assumptions, the model reproduces the zero- and finite-field spectra of all three samples, supporting a common localized-relaxation framework.

\begin{figure}[h!]
    \centering
    \includegraphics[width=0.99\linewidth]{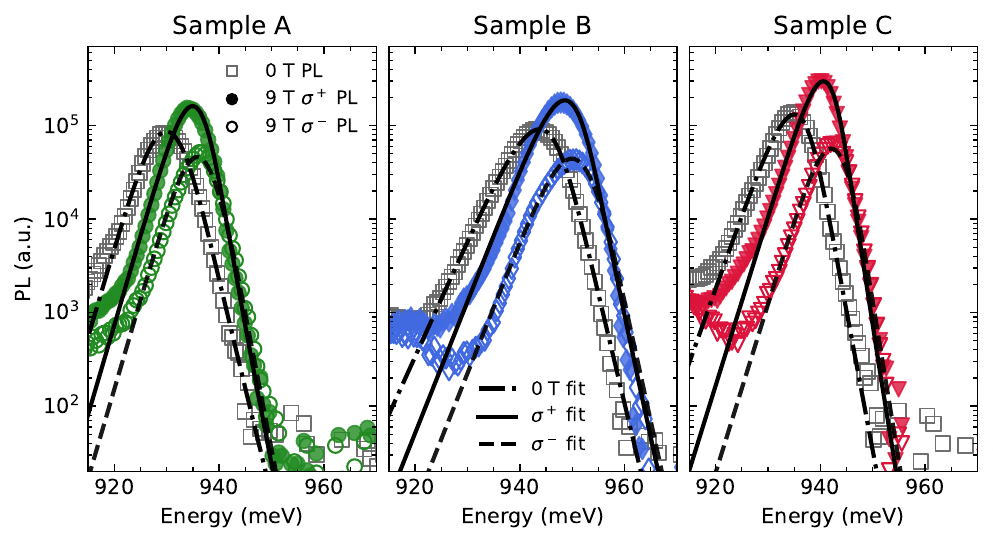}
    \caption{\label{fig:PL_SS_fits}
    Steady-state QW photoluminescence of Samples~A-C at $0$ and $9~\mathrm{T}$, together with fits obtained from the localized-tail rate-equation model. Gray open squares show the zero-field spectra, while filled and open colored symbols show the $\sigma^{+}$ and $\sigma^{-}$ components at $9~\mathrm{T}$, respectively. Dashed, solid, and dash-dotted black curves indicate the corresponding fits to the $0~\mathrm{T}$, $\sigma^{+}$, and $\sigma^{-}$ spectra.
    }
\end{figure}
\begin{figure*}
    \centering
    \includegraphics[width=0.99\linewidth]{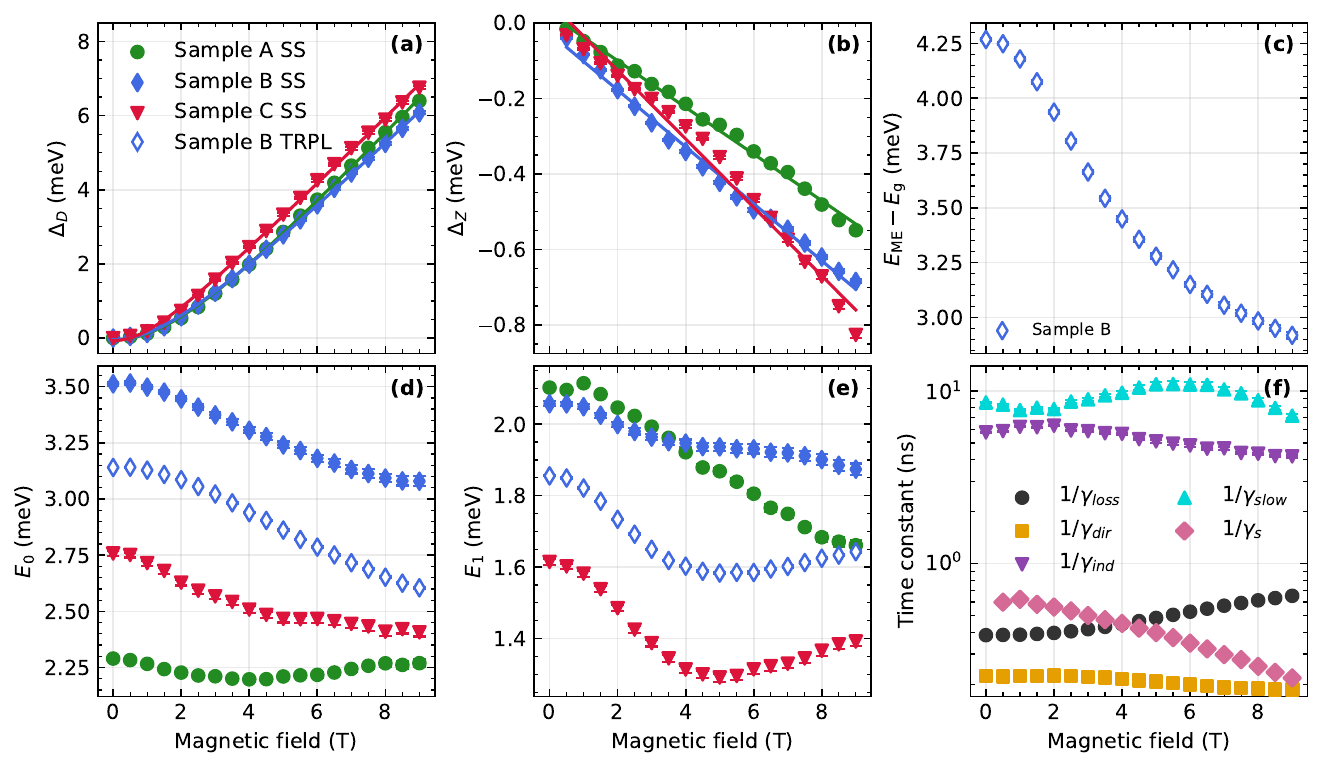}
    \caption{\label{fig:param_b}
    Magnetic-field dependence of the parameters extracted from the localized-state model.
    (a) Diamagnetic shift $\Delta_D$ and (b) Zeeman splitting $\Delta_Z$ obtained from the SS fits of Samples~A-C. Solid lines show the corresponding Fock-Darwin and linear Zeeman fits.
    (c) Field dependence of the kinetic offset $E_{\mathrm{ME}}-E_g$ for Sample~B. Since $\gamma_{\mathrm{hop}}(E_g)$ is fixed, this offset is determined by the fitted $E_1$ and $\gamma_{\mathrm{loss}}$.
    (d) Field dependence of the Urbach energy $E_0$ and (e) hopping-rate energy scale $E_1$.
    (f) Effective time constants extracted from the Sample~B TRPL fits, including local loss, direct feeding, indirect feeding, slow feeding, and spin-flip population exchange.
    }
\end{figure*}

We first consider the orbital response of the localized excitons to the magnetic field. The diamagnetic shift in Figure~\ref{fig:param_b}(a) evolves from an approximately quadratic dependence at low field toward a more linear, Landau-like behavior at higher fields, as expected when magnetic confinement becomes increasingly important \cite{laurindo2022spin}. Assuming an effective parabolic lateral confinement potential, we describe this evolution with the ground-state Fock--Darwin dispersion \cite{llorens2019topology},
\begin{equation}
\Delta_{\mathrm{D}}(B) = \sqrt{ \left(\frac{\hbar^{2}}{m^{*}l_{0}^{2}}\right)^{2} + \left(\frac{\hbar eB}{2m^{*}}\right)^{2} } - \frac{\hbar^{2}}{m^{*}l_{0}^{2}},
\label{eq:fock_darwin}
\end{equation}
where $m^{*}$ is the effective mass and $l_{0}$ the zero-field lateral confinement length. Since $\Delta_{\mathrm{D}}$ is obtained from the field evolution of the fitted $E_g$, it represents an effective ensemble diamagnetic response. The fits yield $l_{0}\simeq22$--$38~\mathrm{nm}$ and $m^{*}=0.059$--$0.066\,m_{0}$, the latter within the $0.052$--$0.077\,m_{0}$ range previously obtained from magnetoluminescence of GaSb/AlGaSb QWs of comparable thickness \cite{giugno1995variation}. The similar scale of the extracted confinement lengths across the three samples indicates broadly comparable lateral localization, while the remaining differences may reflect both the QW structure and the localized-state ensembles weighted by each measurement.

\begin{table}[h!]
    \centering
    \caption{
    Parameters extracted from the magnetic-field dependence of the steady-state energy shifts in Samples A-C. The reduced mass $m^\ast$ and lateral confinement length $l_0$ were obtained from Fock-Darwin fits to the diamagnetic shift $\Delta_{\mathrm{D}}(B)$, while the effective exciton $g$-factor was obtained from linear fits to the Zeeman splitting $\Delta_{\mathrm{Z}}(B)$. Masses are given in units of the free-electron mass $m_0$.
    }
    \label{tab:magnetic_fit_params}
    \begin{tabular}{lcccc}
        \hline\hline
        Sample & Dataset & $m^\ast/m_0$ & $l_0~(\mathrm{nm})$ & $g$ \\
        \hline
        A & SS   & $0.059\pm0.001$ & $22\pm1$ & $-1.07\pm0.02$ \\
        B & SS   & $0.066\pm0.001$ & $24\pm1$ & $-1.30\pm0.03$ \\
        C & SS   & $0.064\pm0.001$ & $29\pm1$ & $-1.56\pm0.05$ \\
        \hline\hline
    \end{tabular}
\end{table}

Figure~\ref{fig:param_b}(d) shows the field dependence of the fitted Urbach energy \(E_0\). At zero field, the steady-state fits give approximately 2.3, 3.5, and 2.8~meV for Samples~A, B, and~C, respectively. With increasing field, \(E_0\) sees a small reduction up to 4~T in Sample~A and nearly recovers its zero-field value by 9~T. In Samples~B and~C instead, the decrease is larger and persists to the highest field. In GaAs/AlGaAs QWs, magnetic compression narrowed the PL line in thicker wells by reducing the contribution of fluctuations on the scale of the exciton radius, but broadened it in thinner wells through reduced averaging of short-range interface disorder \cite{bansal2007magnetic}. Monolayer-scale thickness variations produce larger energy shifts in thinner QWs. Downhill relaxation may thus favor emission from a narrower range of localized states in Sample~A, consistent with its smallest $E_0$. A broader range may contribute in the wider wells, particularly in Sample~B given its non-integer ML nominal thickness.

The evolution of $E_1$ in Figure~\ref{fig:param_b}(e) probes a different aspect of the same localized manifold. Microscopic descriptions of localized-exciton transfer include, besides the density of available final states, the spatial overlap between localization sites and the phonon-assisted transition matrix element \cite{gourdon1989exciton, zimmermann1997excitons}. In the present model, the availability of lower-energy states is represented explicitly by the branching probabilities through $\rho(E)$, whereas their spatial arrangement is not. Toward the low-energy tail, the decreasing density of localized states implies an increasing characteristic separation between localization sites and a stronger role of hopping or tunnelling in the transport \cite{jahn2006influence}. A perpendicular magnetic field can further reduce this connectivity by compressing the exciton wave function, thereby decreasing the overlap between neighboring localized states, while a magnetic-field-induced increase in the exciton translational mass, as commented above, has been shown to suppress lateral exciton motion and was suggested to enhance localization in the presence of potential fluctuations \cite{akiyama1995suppression}. These effects provide a possible microscopic origin for the decrease of $E_1$: transfer toward the more spatially separated and weakly overlapping states deeper in the localized manifold would be increasingly hindered, causing the redistribution rate to fall more rapidly with decreasing energy. The corresponding $E_{\mathrm{ME}}$ in Figure~\ref{fig:param_b}(c) is not an additional free parameter of the field-dependent fits. Since $\gamma_{\mathrm{hop}}(E_g)$ is fixed to its zero-field value, $E_{\mathrm{ME}}-E_g=E_1\ln[\gamma_{\mathrm{loss}}/\gamma_{\mathrm{hop}}(E_g)]$, and its field evolution is therefore determined entirely by the fitted $E_1$ and $\gamma_{\mathrm{loss}}$.

The Zeeman response in Figure~\ref{fig:param_b}(b) probes the spin degree of freedom of the same localized ensemble. Local confinement, strain, symmetry and valence-band mixing can make the exciton $g$-factor sensitive to the particular localized states sampled by the emission \cite{steffan2002exciton,pryor2006lande}. The three steady-state fits yield comparatively similar negative effective $g$-factors, between $-1.07$ and $-1.56$.

Within the coupled-branch model, the polarization imbalance reflects both the Zeeman energy bias and interbranch spin transfer, with $\tau_{\mathrm{s}}$ setting the characteristic time scale of the latter. The values extracted from the polarization-resolved TRPL of Sample~B (Figure~\ref{fig:param_b}(f)) remain between approximately $0.2$ and $0.6~\mathrm{ns}$ and are held fixed in the steady-state analysis. In GaAs/AlGaAs and related III--V QWs, exchange-mediated bright-exciton depolarization occurs on $10$--$100~\mathrm{ps}$ time scales, while the underlying electron- and hole-spin channels extend from a few picoseconds to several hundred picoseconds or longer \cite{maialle1993exciton, munoz1995spin, amand2008exciton}. Localization can shift these scales substantially by restricting centre-of-mass motion and modifying momentum scattering, wave-function overlap and the exchange interactions governing spin relaxation \cite{munoz1995spin,lu2008localized, fokina2010spin, dirnberger2019ultralong}. The underestimated low-energy $\sigma^{-}$ population at $9~\mathrm{T}$ in Figure~\ref{fig:trpl_magneto_fit}(b),(d) is therefore compatible with a reduction of interbranch transfer toward the more strongly localized part of the manifold. An energy-independent $\tau_{\mathrm{s}}$ may consequently transfer population too efficiently at the lowest energies.

Finally, the proximity of GaSb/AlGaSb QWs to the \(\Gamma\)--L conduction-band crossover provides another possible contribution to the unresolved slow dynamics. Quantum confinement can reverse the energetic ordering of the two valleys, producing a thickness-dependent transition between direct- and indirect-gap structures \cite{cebulla1987direct,chellu2021highly}, while magnetic fields can further modify their relative energies \cite{forchel1987temperature}. If local well-width fluctuations bring parts of the present QWs close to this crossover, states with increased L-valley character could provide an additional population reservoir or transfer pathway, potentially contributing to the weak non-monoexponential component represented phenomenologically by the indirect-feeding channel. The present measurements cannot distinguish such a contribution from the other mechanisms discussed above.
\section{Conclusion} \label{sec:conclusion}

We have shown that disorder-mediated energy relaxation and magnetic-field-dependent spin dynamics jointly govern the optical response of the QWs surrounding local-droplet-etched (In)GaSb/AlGaSb QDs. By combining energy- and polarization-resolved measurements with a localized-tail rate-equation model, we distinguish downhill relaxation through localized exciton states from the population redistribution induced by spin-flip scattering. The model reproduces the complete spectral and temporal evolution at zero field and, when extended to coupled Zeeman branches, accounts for the pronounced differences between their polarization-resolved dynamics. We extract effective exciton $g$-factors ranging from $-1.07$ to $-1.56$ and spin-flip times of $0.2$-$0.6~\mathrm{ns}$. The agreement with steady-state spectra across different well structures, using kinetic parameters constrained by time-resolved measurements, further supports this common relaxation framework.

Our results also reveal the limits of this effective description. In particular, the persistence of low-energy emission in the weaker Zeeman branch suggests that spin relaxation may become less efficient in more strongly localized states. Overall, the measurements establish a temporal separation between the dominant QW response and the development of QD emission, confirming the relevance of the surrounding well as an intermediate carrier reservoir. Further measurements varying temperature and excitation conditions, including resonant and quasi-resonant excitation with opposite polarizations, could clarify the mechanisms governing carrier relaxation and capture. Establishing how these processes affect QD occupation and secondary carrier capture would help identify strategies for improving single-photon purity and preserving spin coherence.

\section*{Acknowledgements}

This publication is part of the R\&D\&I project PID2022-136626OB-C33, funded by MICIU/AEI/10.13039/501100011033 and by ERDF, EU. We also acknowledge financial support from the European Union's Horizon 2020 research and innovation programme under the Marie Sk\l{}odowska-Curie grant agreement No.~956548 (project Quantimony); the CSIC Interdisciplinary Thematic Platform (PTI+) on Quantum Technologies (PTI-QTEP+); the Research Council of Finland projects QuantSi (Decision No.~323989) and CryoLight (Decision No.~357351) and Flagship Programme PREIN; and Business Finland, through the co-innovation project QuTI (41739/31/2020) and the Rise to Challenge project TeleQuant (1835/31/2025).

\bibliographystyle{unsrt}
\bibliography{sample}

\end{document}